# Defense Priorities in the Open-Source AI Debate

## A Preliminary Assessment

*By Masao Dahlgren*                                           AUGUST 2024

## THE ISSUE

- **A spirited debate is taking place over the regulation of *open foundation models***—artificial intelligence models whose underlying architectures and parameters are made public and can be inspected, modified, and run by end users.
- **Proposed limits on releasing open foundation models may have significant defense industrial impacts**. If model training is a form of defense production, these impacts deserve further scrutiny.
- **Preliminary evidence suggests that an open foundation model ecosystem could benefit the U.S. Department of Defense's supplier diversity, sustainment, cybersecurity, and innovation priorities.** Follow-on analyses should quantify impacts on acquisition cost and supply chain security.

---

Generative artificial intelligence (AI) has increasingly become a U.S. Department of Defense (DOD) priority. Software powered by generative AI *foundation models*—generalist systems that emulate human reasoning—might process reams of raw intelligence, automate Pentagon paperwork, or allow aircraft, trucks, and ships to navigate themselves.[1] Many advancements in this sector originate in commercial and academic research.[2] If generative AI sees wide adoption across the DOD, this base of commercial foundation model developers will become a critical part of the defense industrial base.[3] The Joint Force thus has a stake in the commercial foundation model ecosystem and how it evolves.

Indeed, DOD AI strategies hinge on continued commercial innovation in AI.[4] To that end, the Pentagon has assigned new funding to acquire AI-powered systems, such as for its Replicator drones and Joint All-Domain Command and Control battle network, and new organizations to manage them, empowering the Chief Digital and AI Office (CDAO), Task Force Lima, and others.[5]

Amid this institutional buildup, the Pentagon should appraise proposed commercial foundation model market regulations. As with spectrum auctions or shipbuilding, civil sector policymaking will shape the DOD's future choices.[6] Policies that create a competitive ecosystem of market players could improve the supply chain for future DOD programs. Conversely, policies that accelerate consolidation—like the Jones Act or the 1993 "Last Supper"—might threaten it.[7] The 2023 Executive Order on the Safe, Secure, and Trustworthy Development and Use of Artificial Intelligence has made unprecedented use of the Defense Production Act to regulate the commercial AI market.[8] It is now worth asking where regulation of commercial markets could affect defense production.

**Figure 1: Example of 1990s-Era Defense Industrial Consolidation**

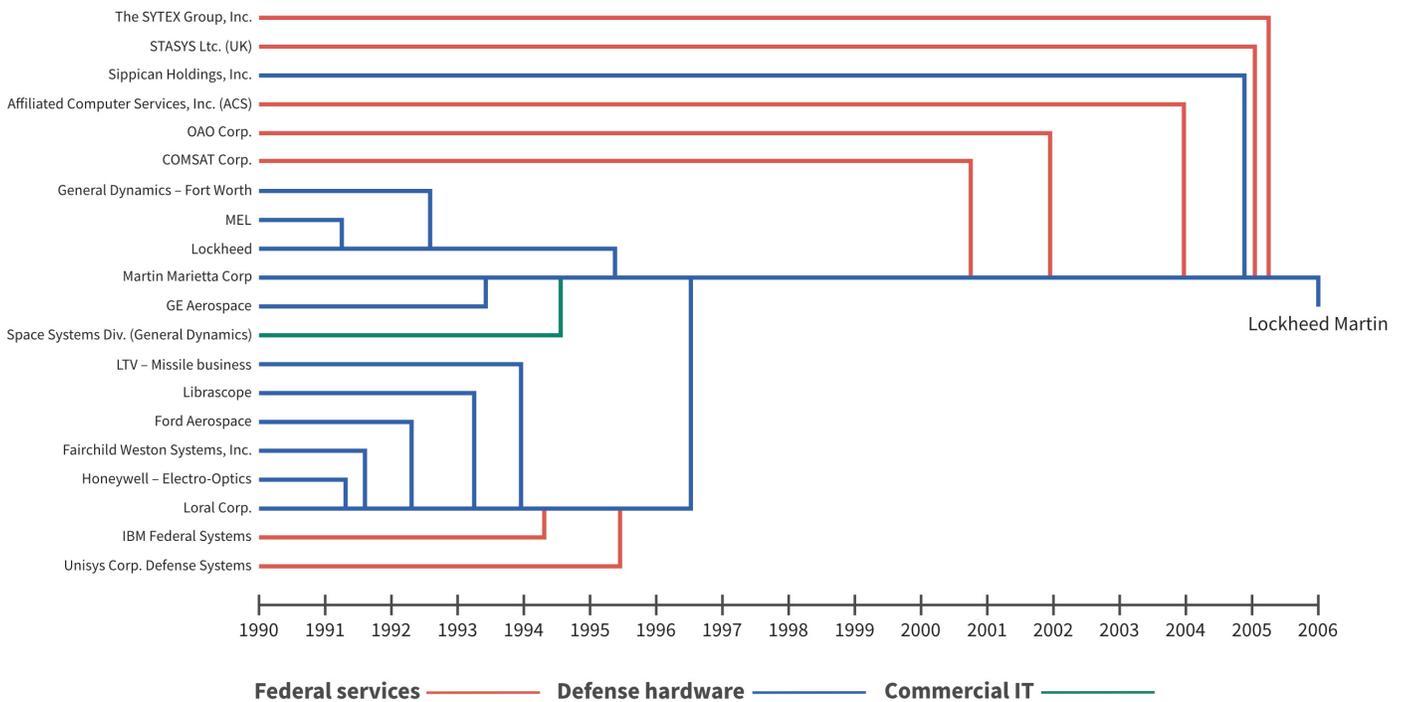

Source: CSIS.[9]

## THE OPEN FOUNDATION MODEL DEBATE: A LITMUS TEST

Emerging civil society debates over AI safety–especially over open foundation models–merit particular attention. Unlike with closed models like GPT-4, developers of open foundation models like Llama, Mistral, or Qwen openly publish the models' underlying parameters ("weights"), allowing them to be inspected, modified, and operated by end users.[10] With the performance of open models approaching their closed counterparts (Figure 2), some have suggested that open model distribution could pose "extreme risks" for misuse.[11] Others, meanwhile, have highlighted open models' benefits for research, security, and national competitiveness.[12] Though outcomes remain uncertain, proposals to limit the distribution of open models–such as through California Senate Bill (SB) 1047–have recently gained legislative traction.[13]

How the open foundation model debate is resolved would have direct implications for the defense industrial base. As detailed in later sections, there are preliminary reasons to believe that a diverse open model ecosystem might benefit the DOD. The widespread availability of high-performance, open-source foundation models could improve the DOD's ability to (1) competitively source and sustain AI systems, (2) deploy AI securely, and (3) address novel use cases. Considering these impacts, the open model debate represents a test case for how civil society evaluates defense priorities in AI policy decisions.

Outlining these implications might also clarify, in White House Office of Science and Technology Policy director Arati Prabhakar's words, an often "garbled conversation about the implications, including safety implications, of AI technology."[14] Indeed, in its flagship report on the subject, the Biden administration suggested that "the government should not restrict the wide availability of model weights" but that "extrapolation based on current capabilities and limitations is too difficult to conclude whether open foundation models, overall, pose more marginal risks than benefits."[15] The administration has not endorsed open model restrictions nor foreclosed future regulation. An accounting of defense industrial benefits might therefore contribute to this ongoing conversation.

## TERMS OF THE DEBATE

Open-source software and standards are already widespread in U.S. national security applications.[16] Army smartphones, Navy warships, and Space Force missile-warning satellites run on Linux-derived operating systems.[17] AI-powered F-16s run on open-source orchestration frameworks like Kubernetes, which is regularly updated, maintained, and tested by industry and the broader public.[18] Open-source software is ubiquitous, permeating over 96 percent of civil and military codebases, and will remain a core piece of defense infrastructure for years to come.[19]

What constitutes an "open" foundation model is less well defined. Developers can distribute foundation models at different levels of "openness"—from publishing white papers and basic technical information to releasing models entirely, including their underlying weights, training data, and the code used to run them.[20] By contrast, developers of closed models, including GPT-4 or Claude, release fewer details or data, only allowing user access through proprietary application programming interfaces.[21] In general, this brief defines "open" models as those with widely available weights, consistent with relevant categories in the 2023 AI executive order.[22] Many of the risks and benefits discussed here flow from these definitions.

Claims of extraordinary risk have motivated several recent proposals surrounding open-source AI. Analysts have expressed concern that malicious users might modify open foundation models to discover cybersecurity vulnerabilities or instruct users in the creation of chemical and biological weapons.[23] Others have argued that public distribution of model weights could aid adversaries in advancing their AI capabilities.[24] Given these apprehensions, some observers have proposed export controls, licensing rules, and liability regimes that would limit the distribution of open foundation models.[25]

A competing school of thought has emphasized the societal benefits of open foundation models.[26] Open distribution of weights, some argue, accelerates innovation and adoption: indeed, the key frameworks and innovations underpinning today's large language models (LLMs), like PyTorch and the transformer architecture itself, were distributed openly.[27] Others contend that the public scrutiny of model weights enables rapid discovery and repair of vulnerabilities, improves public transparency, and reduces the concentration of political and economic power as AI systems increase in importance.[28]

What is most clear, however, is that this risk-benefit assessment remains incomplete. The U.S. Department of Commerce's initial assessment is inconclusive, and AI safety literature has thus far lacked clear frameworks for identifying relative risk and benefit and whether they are unique to open models.[29] Despite concerns over AI models instructing untrained users in biological weapon development, for instance, recent red-teaming exercises concluded that LLM-equipped teams performed similarly to those without.[30] Similar concerns over AI-assisted cyber vulnerability discovery remain unclear, with some arguing that enhanced vulnerability detection may benefit cyber defenders over attackers, or that the balance of advantage would be case-dependent.[31] Malicious use, meanwhile, continues to take place with closed models.[32] In brief, more research remains necessary to unpack where the relative risks and benefits lie.[33] The purportedly catastrophic harms of tomorrow's foundation models have not yet come into clear view.[34]

Second, the pace of technical change has been so uncertain that evaluating future benefits, harms, and policy interventions can be challenging.[35] Whether a licensing regime is effective, for example, depends on how readily foundation model technologies will diffuse.[36] And whether export controls benefit national security hinges on which analogy becomes relevant: Is restricting open models like restricting nuclear weapons exports, or is it akin to Cold War bans (now repealed) on public-key cryptography, a technology which now underpins online banking, e-commerce, and a multi-trillion-dollar digital economy?[37] In the absence of a U.S. market presence, will Chinese open models take their place?[38]

Finally, questions remain on how to implement AI policy. Definitional challenges abound; early AI policy approaches, including the EU AI Act, AI executive order, and California SB 1047, apply thresholds for "systemic risk" to models exceeding a certain amount of computing power or cost used in their development.[39] However, such thresholds for triggering government review, such as the $10^{26}$ floating-point-operation threshold in the AI executive order, may incompletely capture the capabilities they aim to regulate.[40] How to balance resourcing for AI policy implementation against other cyber and biological threat mitigations, such as for material monitoring or new cyberdefense capabilities, remains another open question.[41]

## Figure 2: Performance Comparison of Large Open-Weight and Closed Models as of July 2024

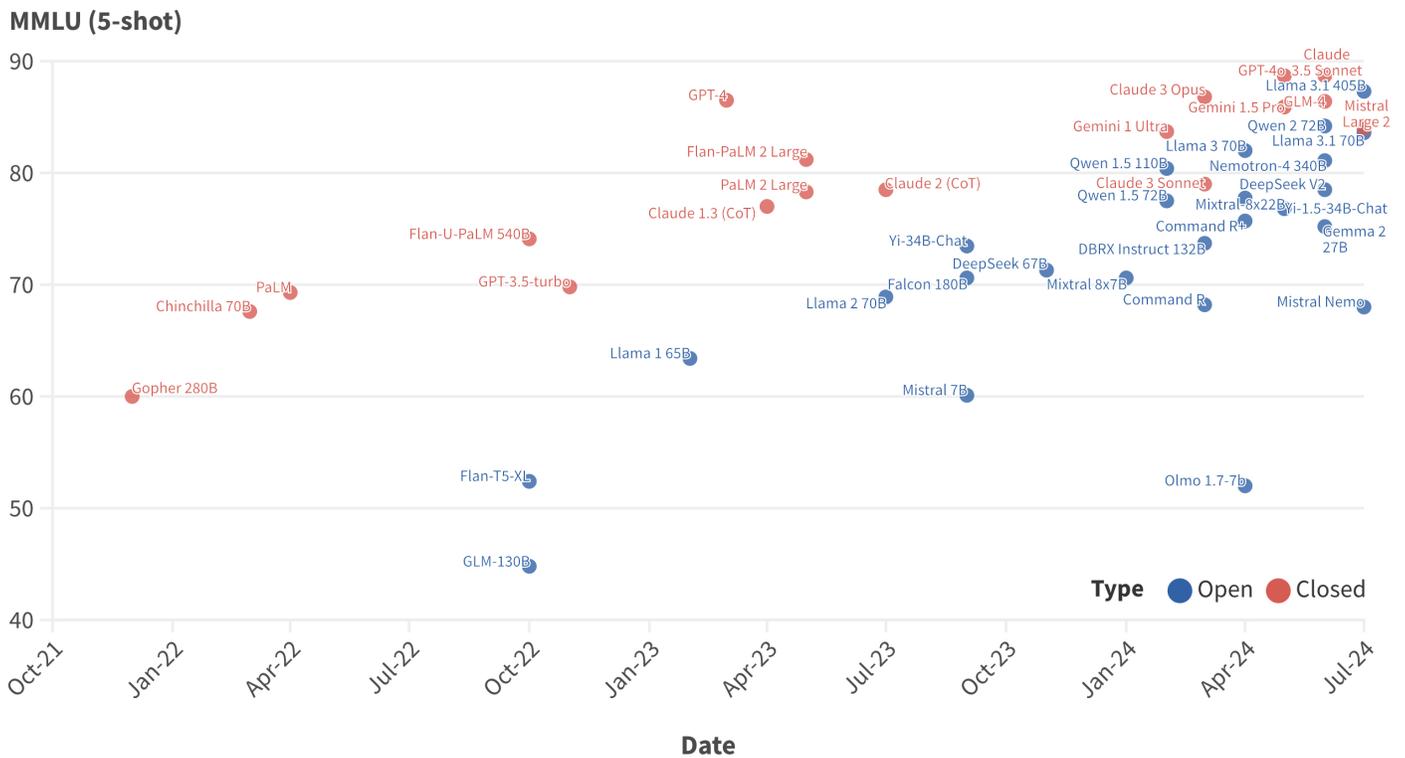

Source: CSIS (see appendix). Chart inspired by Maxime Labonne.[42]

## OPEN SOURCE AND THE JOINT FORCE

A defense industrial assessment could thus contribute a valuable perspective to the AI risk debate. With AI industry trends favoring consolidation, the open foundation model ecosystem may become an increasingly important source of competition in the industrial base.[43] Because end users can modify and run open models directly, they have become increasingly relevant for developing local, secure applications and embedded systems—needed by military users demanding low power usage, security, and reliability. And because open models can be publicly inspected, red teamed, and verified, they may present defense-related cybersecurity advantages.[44]

To date, however, the DOD has largely focused on AI adoption.[45] In its flagship data, responsible AI, and adoption strategies, the DOD has focused on harnessing private sector innovations for national security end uses.[46] It has embedded chief data officers in combatant commands; tested AI use cases in major experimentation initiatives, like the Global Information Dominance Experiment, Proj-ect Convergence, and others; and developed the Responsible AI Framework, emphasizing the use of traceable, transparent AI systems.[47]

In August 2023, the DOD established Task Force Lima, an element within the CDAO tasked with "responsibly pursu[ing] the adoption" of generalist models.[48] Alongside the CDAO and Responsible AI Working Council, Lima was chartered to "accelerate" AI initiatives, "federate disparate developmental and research efforts," and engage across the interagency on the "responsible development and use of generative AI," with a final strategy due for release in early 2025.[49]

Clarifying potential AI use cases within the DOD is a valuable first step in "mak[ing] life easier for program offices who want to do AI or add AI."[50] A valuable second step would be to identify the broader trajectory of the AI industrial base. The DOD will often rely on industry expertise to develop and identify more generative AI use cases; a broad ecosystem of model and application developers will be critical for this process.[51]

In short, an assessment of defense industrial impacts is conspicuously missing from the broader debate on open foundation models. Arguments over model regulation are couched in national security language but should involve a broader swath of national security practitioners, including defense acquisition professionals.[52] Accordingly, DOD elements, including the CDAO, should independently assess the national capacity to develop AI-powered systems and the impact of open foundation models.

*Arguments over model regulation are couched in national security language but should involve a broader swath of national security practitioners.*

## FROM ADOPTION STRATEGIES TO INDUSTRIAL STRATEGIES

A defense industrial accounting is needed because of preliminary evidence that open foundation models–and their supporting ecosystem–could be useful for the DOD. AI adoption remains a DOD priority, and open release has historically accelerated the rate of technology adoption.[53] Open-source development may have positive competitive implications for defense acquisition. And open-source communities are accelerating developments in on-premises deployment, fine-tuning for specialized applications, model reliability, and other desirable characteristics for defense end users.[54]

1. **Supplier diversity.** In its 2023 National Defense Industrial Strategy (NDIS), the DOD prioritized the diversification of national component supplier bases.[55] To improve competition for new entrants, for instance, it has developed new funding mechanisms, like Strategic Funding Increase contracts, and innovation organizations, like the Defense Innovation Unit.[56] A robust ecosystem of open foundation model developers might improve defense supplier diversity and prevent market consolidation.

   There are strong incentives for consolidation in the foundation model industry.[57] Unlike other software products, large foundation models are particularly capital-intensive to develop, rivaling the supply chain complexity of major defense hardware. Training a foundation model, open or closed, demands extremely large datasets, costly graphics processing units and accelerator chips, and talent, advantaging larger players.[58] Operating them–a process termed *inference*–also imposes high energy costs.[59] Due to these factors, industry leaders have spent tens of billions of dollars to develop and deploy so-called frontier foundation models.[60] This high cost of entry increases the risk of industry consolidation and, consequently, capacity pressure on future DOD acquisition programs.

   Would a robust open AI ecosystem relieve these pressures, creating competition for proprietary vendors and accelerating innovation?[61] Open foundation models have begun to show competitive performance with closed, proprietary alternatives.[62] To date, many competitive open foundation models have been developed by well-resourced actors.[63] But community modification and experimentation have sped the development of new architectures and reductions in training and inference costs.[64] Modified open models have also demonstrated high performance in highly specialized tasks; they might be similarly performant for those anticipated for defense applications, like flight control, business automation, or intelligence fusion.[65]

2. **Competitive sustainment.** More critically, the existence of open foundation models might mitigate dependence on single vendors when sustaining AI-powered defense systems.[66] Foundation models used in defense applications may need specialized retuning, including with government data, as the operating environment changes.[67] High-performance closed models have also shown performance drifts over time; access to model weights and other information would help vendors diagnose emergent problems.[68] In short, by building on open foundation models, the DOD could reduce barriers for vendors to compete on model integration, retuning, and sustainment.

   These imperatives resemble those motivating the Pentagon's Modular Open Systems Approach (MOSA) for major hardware purchases.[69] Lack of access to technical data has historically challenged the DOD's ability to reliably and affordably sustain military hardware.[70] Given these barriers, military services have increasingly demanded access to the data needed to service new helicopters, trucks, and ships. For its nearly $70 billion Future Long-Range Assault Aircraft program, for example, the U.S. Army rejected a $3.6 billion lower bid over incomplete MOSA requirements, opting for a vendor that provided full access to aircraft technical data packages (TDPs) required for maintenance.[71] Just as it is challenging to competitively source aircraft maintenance without access to TDPs, competitively sustaining AI-powered software may be difficult without access to model weights.[72]

It would be challenging, however, to adopt such a modular approach with closed-model-powered software. Closed model developers are unlikely to relinquish access to model weights for other vendors to modify while performing their sustainment contracts.[73] Moreover, to securely implement model-level changes to a defense application, national security customers would potentially need to certify both the application vendor and closed model vendor separately, adding additional friction to an already difficult acquisition process.[74] Evaluating appropriate data rights for closed models could demand a further institutional lift.[75]

A world with competitive open foundation models would allow the Pentagon to sidestep these sustainment challenges. While vendors would tune and adapt open models for proprietary applications, visibility into their underlying weights and architectures would make it easier for others to maintain and integrate them, potentially reducing the cost of sustainment.[76] The presence of open options, moreover, could improve taxpayer bargaining power in negotiations with closed providers on complex data-rights matters. The DOD has already invested heavily in competitively sustaining military hardware. Those same lessons apply when sustaining AI-powered software.

3. **Security and reliability.** Concerns over the confidentiality and reliability of foundation models remain key barriers to DOD generative AI adoption.[77] The potential to locally operate open foundation models on DOD infrastructure therefore becomes a key advantage. Open models present useful options for building AI-powered systems without needing to certify an external foundation model developer for sandboxed deployments.[78] Like with Linux, Kuber-netes, or other open software libraries, these models might become a secure baseline for AI that vendors modify with classified or specialized information.[79]

A robust open research community could also drive advancements in AI model reliability and interpretability, reducing the number of *hallucinations*–nonuniform responses that do not reflect the information a user needs.[80] Access to model weights is often crucial for diagnosing failure and evaluating model reliability in detail.[81] Insofar as hallucinations remain a "showstopper" for defense AI adoption, innovations in the open foundation model ecosystem are worth considering.[82]

Finally, much like with open-source software, open access to weights might enable a greater possibility of detecting vulnerabilities.[83] Indeed, the AI executive order places special emphasis on red teaming future foundation models; more open approaches to model publication allow a wider peer review of model performance.[84] As Cybersecurity and Infrastructure Security Agency analysts have recently emphasized, there is "significant value in open foundation models to help strengthen cybersecurity, increase competition, and promote innovation."[85] Accordingly, future assessments might review the advantages and disadvantages open models present for hardened, defense-critical AI systems.

4. **Specialized use cases.** Lastly, a robust open foundation model ecosystem might enable AI use cases that receive less attention from closed-source providers. Because open models can be retrained, fine-tuned, and broadly customized, they can serve as a basis for national-security-specific applications.[86] Further, open-source initiatives can drive innovation in national-security-relevant topics, such as for document and data search with semantic retrieval.[87] These search algorithms leverage embedding models–a form of language model–to compare the semantic meaning of stored documents and return relevant results; open models largely dominate performance metrics for embedding generation.[88] Domains overlooked by major commercial players could benefit from the dynamism of open development.[89] Further investigation should assess use cases where open model performance meaningfully exceeds closed alternatives.

# AN AI "LAST SUPPER"?

Preliminary evidence suggests that open foundation models might benefit the defense industrial base. What is now needed is a quantitative assessment of the open ecosystem's fiscal impacts. Beyond assessing pathways to adoption, defense policymakers should review the changing competitive landscape for foundation models and potential implications for the defense industrial base. Three recommendations follow:

1. **In its forthcoming review of DOD foundation model adoption, Task Force Lima and partners should compare open and closed models in generative AI use cases.** Different foundation model architectures have roles to play in defense applications. For document summarization tasks, for instance, developers might use open- and closed-source models in a variety of architectures, from retrieval-augmented generation to in-context learning. Although technical details on performance are beyond this paper's scope, a broad comparison of potential open- and closed-model use cases could provide useful context for acquisition professionals.

2. **NDIS implementation should include assessments of the foundation model industrial base, including data, infrastructure, human capital, competitive dynamics, and the impact of open model development.** The 2022 National Defense Strategy acknowledges that the DOD "will be a fast-follower where market forces are driving commercialization of . . . trusted artificial intelligence," while the NDIS asserts that it must "diversify [the] supplier base" and "foster competition within the defense market."[90] An independent assessment of those market forces, including competition in the AI market, would be critical to implementing the NDIS.

3. **The DOD should collect data on where open foundation models are used in its systems, the number of nontraditional performers who leverage open foundation models, and models' potential fiscal impacts.** As the DOD begins to adopt AI-powered systems, it should collect data on where open foundation models are being leveraged. Such data might inform future analyses on the health of the AI industrial base and the ability for new entrants to compete.

There is a considerable risk in disregarding defense industrial impacts in the debate over open foundation models. Major consolidation in other sectors, such as in shipbuilding and aerospace, has come alongside major declines in defense acquisition speed and capacity.[91] If open foundation models are indeed "dual-use"—and therefore critical to national security—the potential for consolidation deserves national security attention.

Civil sector policy decisions have created a shipbuilding base outproduced by China 230 to 1.[92] Other deliberate choices over the nation's industrial base have meant that demand now outpaces supply for missile defense systems, artillery shells, and AI talent.[93] If production is deterrence, these are the stakes of the open-source AI debate.[94]

Others will not wait for the United States if it falls back.[95] China and other states have made vast investments in stimulating their domestic AI industries, with top models "not far behind" Western counterparts.[96] The United States' competitors view open foundation models as a means of capturing global market share and advancing scientific and economic development.[97] While the development of AI is not an arms race, it is a broader economic and social competition—one where U.S. priorities on democracy, transparency, and security should define global standards.[98] The technology and value system to do so are already in place. Attention is all it needs.[99] ∎

*Masao Dahlgren is a fellow with the Missile Defense Project at the Center for Strategic and International Studies in Washington, D.C.*

*The project team held two private roundtables, on July 10 and August 1, 2024, to inform the conclusions of this brief. Participants included subject matter experts from the U.S. Department of Defense, U.S. Department of Energy, AI and defense firms, investors, universities, nonprofits, and other stakeholder groups. In addition, the author interviewed multiple government and industry experts to support the research process.*

*Special thanks to Patrycja Bazylczyk for her assistance in roundtable facilitation, report review, and project performance. Further acknowledgments to Shaan Shaikh, Wes Rumbaugh, Tom Karako, Cynthia Cook, Greg Sanders, Charles Yang, Kevin Li, Michelle Fang, and the many external reviewers involved in refining this report.*

*This brief was made possible with support from the Omidyar Network and general support to the CSIS Missile Defense Project.*

# APPENDIX: SELECTED MODEL PERFORMANCE EVALUATIONS

The following table compiles self-reported Massive Multitask Language Understanding (MMLU) scores, one of many benchmarks used to evaluate foundation model performance. Typical scores reported are "5-shot," where five examples are provided before models are prompted with a question. "Open" models listed include those with publicly available model weights.

While MMLU is an incomplete representation of model performance, this benchmark was selected because it is among the oldest, allowing for consistent comparison over time. Writ large, the foundation model industry increasingly suffers from a benchmarking crisis, facing issues of model overfitting–internalizing existing benchmark results–and obsolescence.[100] Developing trusted benchmarks has thus become a major Department of Commerce priority.[101]

| Date | Type | Country | Developer | Model | MMLU 5-shot[102] |
|------|------|---------|-----------|-------|------------------|
| May 2024 | Open | China | 01.AI | Yi-1.5-34B-Chat | 76.8[103] |
| Sep. 2023 | Open | China | 01.AI | Yi-34B-Chat | 73.46[104] |
| Feb. 2024 | Open | China | Alibaba | Qwen 1.5 110B | 80.4[105] |
| Feb. 2024 | Open | China | Alibaba | Qwen 1.5 72B | 77.5[106] |
| June 2024 | Open | China | Alibaba | Qwen 2 72B (base) | 84.2[107] |
| Apr. 2024 | Open | United States | Allen Institute | OLMo 1.7-7B | 52[108] |
| Apr. 2023 | Closed | United States | Anthropic | Claude 1.3 (CoT) | 77[109] |
| July 2023 | Closed | United States | Anthropic | Claude 2 (CoT) | 78.5[110] |
| Mar. 2024 | Closed | United States | Anthropic | Claude 3 Opus | 86.8[111] |
| Mar. 2024 | Closed | United States | Anthropic | Claude 3 Sonnet | 79[112] |
| June 2024 | Closed | United States | Anthropic | Claude 3.5 Sonnet | 88.7[113] |
| Mar. 2024 | Open | Canada | Cohere | Command R | 68.2[114] |
| Apr. 2024 | Open | Canada | Cohere | Command R+ | 75.7[115] |
| Mar. 2024 | Open | United States | Databricks | DBRX Instruct 132B | 73.7[116] |
| Nov. 2023 | Open | China | DeepSeek | DeepSeek 67B | 71.3[117] |
| June 2024 | Open | China | DeepSeek | DeepSeek V2 | 78.5[118] |
| Apr. 2023 | Open | United States | EleutherAI | Pythia 12B | 26.76[119] |
| Mar. 2022 | Closed | United States | Google | Chinchilla 70B | 67.6[120] |
| May 2023 | Closed | United States | Google | Flan-PaLM 2 Large | 81.2[121] |
| Oct. 2022 | Open | United States | Google | Flan-T5-XL | 52.4[122] |
| Oct. 2022 | Closed | United States | Google | Flan-U-PaLM 540B | 74.1[123] |
| May 2024 | Closed | United States | Google | Gemini 1.5 Pro | 85.9[124] |
| Feb. 2024 | Closed | United States | Google | Gemini 1 Ultra | 83.7[125] |

| Date | Type | Country | Developer | Model | MMLU 5-shot[102] |
|------|------|---------|-----------|-------|------------------|
| June 2024 | Open | United States | Google | Gemma 2 27B | 75.2[126] |
| Dec. 2021 | Closed | United States | Google | Gopher 280B | 60[127] |
| Apr. 2022 | Closed | United States | Google | PaLM | 69.3[128] |
| May 2023 | Closed | United States | Google | PaLM 2 Large | 78.3[129] |
| July 2023 | Open | United States | Meta | Llama 2 70B | 68.9[130] |
| Apr. 2024 | Open | United States | Meta | Llama 3 70B (base) | 79.5[131] |
| Apr. 2024 | Open | United States | Meta | Llama 3 70B (instruct) | 82[132] |
| July 2024 | Open | United States | Meta | Llama 3.1 405B (base) | 85.2[133] |
| July 2024 | Open | United States | Meta | Llama 3.1 70B (base) | 79.3[134] |
| July 2024 | Open | United States | Meta | Llama 3.1 405B (instruct) | 87.3[135] |
| July 2024 | Open | United States | Meta | Llama 3.1 70B (instruct) | 83.6[136] |
| Feb. 2023 | Open | United States | Meta | Llama 1 65B | 63.4[137] |
| Sept. 2023 | Open | France | Mistral AI | Mistral 7B | 60.1[138] |
| July 2024 | Closed | France | Mistral AI | Mistral Large 2 | 84[139] |
| Jan. 2024 | Open | France | Mistral AI | Mixtral 8x7B | 70.6[140] |
| Apr. 2024 | Open | France | Mistral AI | Mixtral-8x22B | 77.75[141] |
| July 2024 | Open | France, United States | Mistral AI, NVIDIA | Mistral Nemo | 68[142] |
| June 2024 | Open | United States | NVIDIA | Nemotron-4 340B | 81.1[143] |
| May 2020 | Closed | United States | OpenAI | GPT-3 | 43.9[144] |
| Nov. 2022 | Closed | United States | OpenAI | GPT-3.5-turbo | 69.8[145] |
| Mar. 2023 | Closed | United States | OpenAI | GPT-4 | 86.5[146] |
| May 2024 | Closed | United States | OpenAI | GPT-4o | 88.7[147] |
| Sept. 2023 | Open | United Arab Emirates | Technology Innovation Institute UAE | Falcon 180B | 70.6[148] |
| Oct. 2022 | Open | China | Zhipu AI | GLM-130B | 44.8[149] |
| June 2024 | Closed | China | Zhipu AI | GLM-4 | 86.4[150] |

# ENDNOTES